\documentclass{amsart}
\usepackage{amsfonts}
\usepackage{amssymb}
\usepackage{amsmath}
\usepackage{color}
\usepackage{graphicx}
\usepackage{geometry}
\usepackage{neuralnetwork}
\usepackage{tikz,pgfplots,pgf}
\usetikzlibrary{matrix,shapes,arrows,positioning}

\def \R {{\mathbb {R}}}
\def \N {{\mathbb N}}
\def \C {{\mathbb C}}

\begin{document}

\title[Cortical architectures as contact and sub-Riemannian geometry]
{ Cortical Functional architectures \\ as contact and sub-Riemannian geometry}

\author{Giovanna Citti}
\author{Alessandro Sarti}

\thanks{\noindent
Giovanna Citti, Dipartimento di Matematica, University of Bologna  \\ Alessandro Sarti, CNRS-CREA, Ecole Polytechnique, Paris } 
\maketitle

%\tableofcontents

\begin{abstract} In the paper \cite{[SCP08]}, Jean Petitot  together with the authors of the present paper describe the functional geometry of the visual cortex as the symplectization of a contact form to describe the family of cells sensitive to position, orientation and scale. In the present paper, as a "homage" to the enormous contribution of Jean Petitot to neurogeometry,  we will extend this approach to more complex functional architectures built as a sequence of contactization or a symplectization process, able to extend the dimension of the space. We will also outline a few examples where a sub-Riemannian lifting is needed. 
%Connectivity kernels are induced by propagation dynamics on the 
%associated contact and symplectic distributions. , and emergence of more and more articulated perceptual units in the different structures are described.
\end{abstract}

\section{Introduction}

The mammalian visual cortex presents a modular structure since it is constituted by many families of cells, each one sensitive to a specific feature of the stimulus image: position, orientation, scale, {color}, curvature, movement, stereo and many others. Each of these families presents a  topographical organization in such a way that over every point $x=(x_1,x_2)$ of the retinal plane there is an entire set of cells, each one sensible to a particular instance of the considered feature. The topographical organization takes the form of a hypercolumnar structure where every hypercolumn represents a point of the retinal plane respecting retinotopy {\cite{[HW77]}}.

	In his celebrated paper {\cite{Hoff}} W.C. Hoffmann introduced for the first time differential fiber bundles for the description of the visual cortex. Here the base of the fiber bundle represents the retinal plane while fibers represent the engrafted variables. After his work, subsequent studies were mainly focused on the set of simple cells responsible for the detection of position and orientation of edges. A similar differential approach based on Elie Cartan moving frames has been adopted by Steven Zucker (see \cite{[Z05]} and the references therein). He noted that differential geometry instruments can be used to formalize the process of boundary completion. 
	
	Jean Petitot, together with Yannick Tondut  \cite{[TP97]} \cite{[PT99]}, described the set of simple cells as a fiber bundle equipped with a contact structure and introduced for the first time the notion of neurogeometry, that is the intrinsic geometry of neural connectivity. They identified the connectivity structure between simple cells with the Heisenberg group, and performed contour completion in this structure minimizing a suitable length functional. Co-axial long range connectivity found by neurophysiological experiments  has been modelled as integral curves of the contact structure. As well as the boundary completion of celebrated Kanizsa triangle has been accomplished as minimization of a functional constrained by the contact structure. 
	
	Based on the important work of Petitot and Tondut \cite{[TP97]} \cite{[PT99]}, the authors of the present paper proposed to represent the functional architecture in terms of stratified Lie groups with a sub-Riemannian metric to take into account the local symmetry of the cortex \cite{[CS06]}. The hypercolumnar structure of simple cells is constituted by the complete set of cells sensitive to all orientations and each cell in the fiber is obtained from a fixed one by rotation. Since each hypercolumn is obtained translating a fixed one, the structure of this layer of cells is identified with the Lie  group of translations and rotations. It is well known that different geometries could be defined on this group. The sub-Riemannian metric was chosen, since the integral curves of its generating vector fields can also be considered as a mathematical representation of the association fields of Field, Hayes and Hess {\cite{FHH}}. 
	In this model contour completion is modelled as a front propagation in the sub-Riemannian geometry  {\cite{[CS06]}, \cite{BD}, \cite{DB}}.  In this structure it has been proved the relation between neural mechanisms and perceptual  completion in both cases of modal and a-modal completion. See \cite{[CS14_2]} for a review.
	Analytical properties of this model were further studied by R. Hladky and Pauls {\cite{[HP08]}}, and a semidiscrete version was proposed in  \cite{PG}. Finally we recall the works of Duits, van Almsick, Franken, ter Haar Romeny  {\cite{FrankenDuits} \cite{[ADFH05]}} who proposed models of image processing in the same Lie group.

As an extension of the models of Petitot \cite{[PT99]} and Citti-Sarti \cite{[CS06]}, the three authors together proposed a model of functional connectivity of simple cells in a symplectic structure  \cite{[SCP08]}, taking into account translation, rotation and scale. The approach has been particularly fecund, since  it reveals the pattern of co-axial and trans-axial connections found by neurophysiological experiments as integral curves of the symplectic structure. The technique adopted to extend the previous models consist to transform the one-form of the contact structure in the two-form of the symplectic one, performing an operation called {\it symplectization}. A similar procedure but considering hyperbolic metrics has been adopted by the same authors in  \cite{[SCP09]}. 
We also quote the following more recent neurogeometric models of the functionality of the cortex 
\cite{Aleksee, Chossat, BCFFP, BHR, MP2015, THR14, KL2018, LM2023}. 
For a more complete introduction to these models,  application to  perceptual phenomena or  image analysis, we refer to \cite{[P14]}, the entire volume \cite{[CS14]} and the references therein.

\begin{figure}[tbh]\label{fig1}
\includegraphics[height=1in]{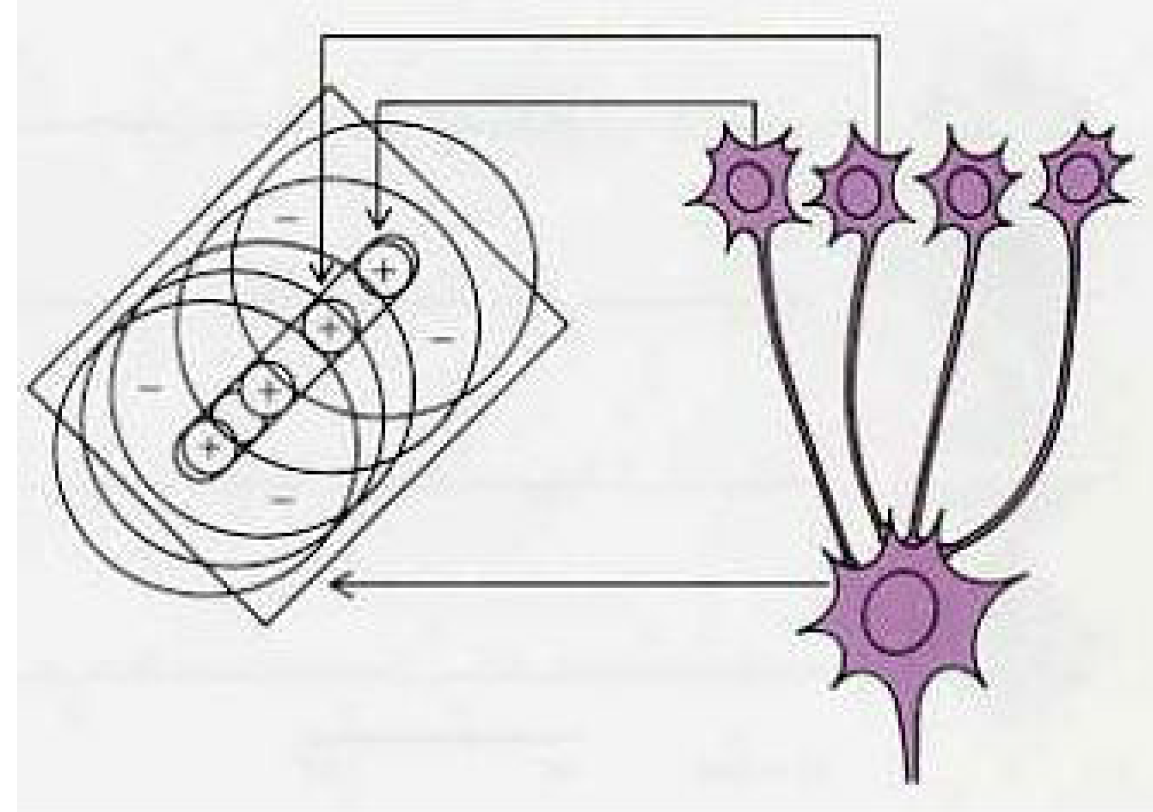}
\includegraphics[height=1.2in]{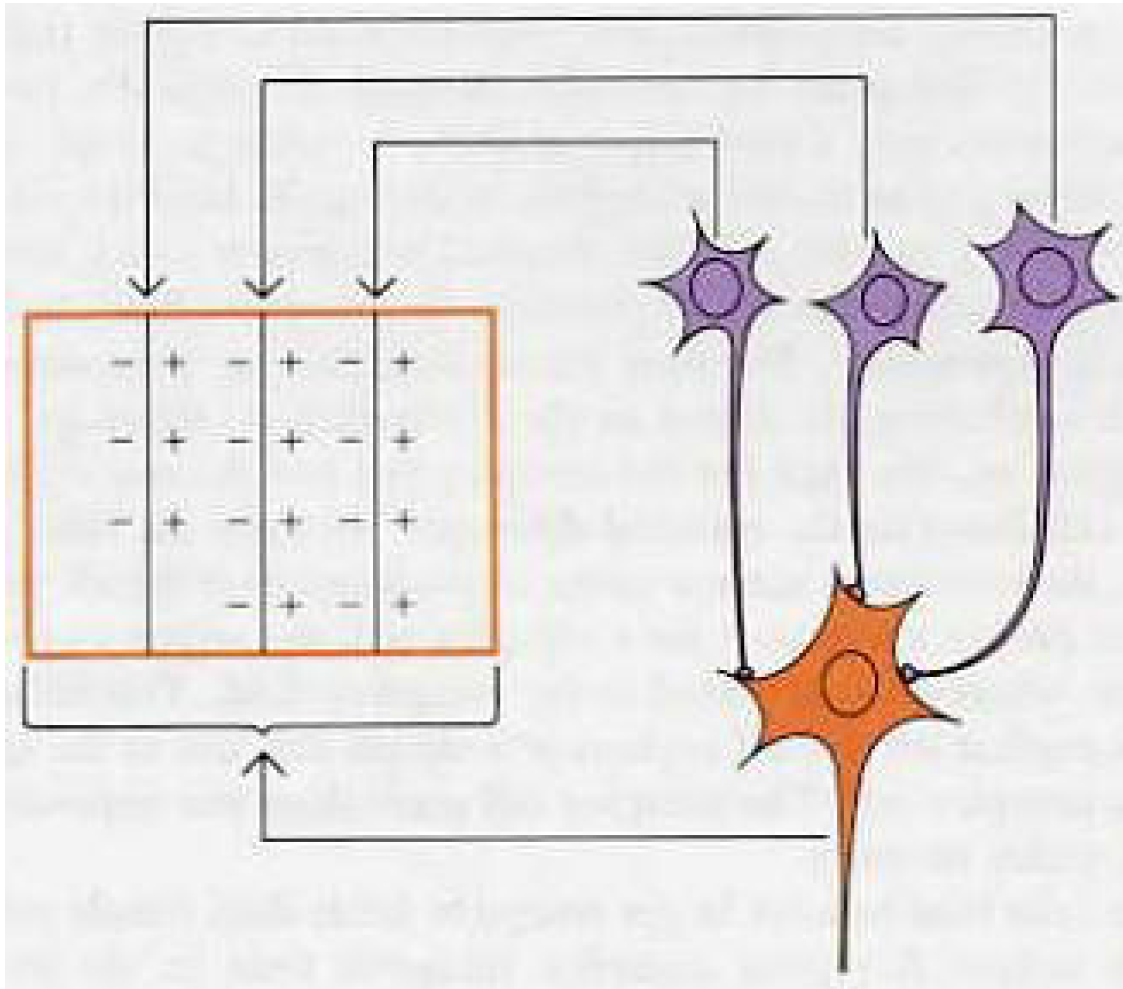}
\caption{The model of Hubel-Wiesel: the connectivity produces a simple cell receptive profile starting from an alignment of retinal cells (left), and produces a complex receptive profile starting from simple cells. 
Image taken from \cite{H}. }
\end{figure}

In the present paper, that we are delighted to dedicate to Jean Petitot as a {\it homage} for his 80th birthday, we will reconsider this approach and will extend it to describe the modularity of the visual cortex. Particularly we will consider subsequent  {\it contactization}  and  {\it symplectization} procedures to model the neurogeometry of different cortical layers and families of cells. 
Our approach is compatible with the model proposed by 
Hubel and Wiesel in \cite{[HW77]}. In fact they  introduced a type of wiring which could produce a simple-cell receptive field starting from an alignment of retinal cells with radially symmetric receptive field (see figure \ref{fig1} left) . 
A similar type of wiring, produces a complex cell, starting from an alignment of simple cells (see  figure \ref{fig1} right). 

Is an open problem to understand if this is just a mathematical procedure or the natural process that has been implemented during evolution to introduce new kind of cells from exiting ones. 

%
%Here we focus on the modularity of the cortex. Starting from the model of Hubel \cite{}, we  show that each family of cells acts in a comparable way on the family of cells from which it takes input, but the result is completely different, since the input is different. 
%In the paper \cite{}, in collaboration with J. Petitot, the authors of the present paper describe simple cells as a contact structure, 
%
%and use a symplectization to describe the family of cells sensitive to scale. 
%
%the action of the connectivity,  
%able to extract new features, as a contactization or a symplectization process, able to extend the dimension of the space. The connectivity kernels will be described as in
%
%In higher areas more abstract process take place, and perceptive units are handled. 
%This will be modelled as a spectral analysis and a quotient of the geometric structure will take place, allowing to geometrically describe the passage. 

\section{From functional geometry to differential structure via contactization,  symplectization and sub-Riermannian lifting}

Contactization and symplectization are well known mathematical procedures, which starting respectively from a symplectic or a contact space provide a contact or a symplectic one, by adding new variables. This instrument was first applied to model the functionality of the visual cortex  by Jean Petitot in \cite{[TP97]},\cite{[PT99]}, and then extended in \cite{[SCP08]} and \cite{[SCP09]} (see also \cite{LM2023}).  However they are insufficient to generate the large variability of the families of cells in V1, and we will also apply sub-Riemannian lifting.

\subsection{The retinal plane}\label{retina}

The retinal plane $M$ is often identified with the $\R^2$ space with an Euclidean geometry.  This representation is an idealization, and it is a good approximation just close to the foveal point because it discards the fact that the retina is a half sphere.  A natural reference frame in this setting is simply induced by a choice of coordinates and the classical choice of vector fields 
$$X_{0; 1}= \partial_{x_1}, \;\;X_{0; 2} = \partial_{x_2}.\footnote{\label{footnote1}For every family of cells, we will define a possibly different choice of vector fields. Hence we use the notation  $X_{h;i}$, where $h$ identifies the current family of cells and  $i$ identifies the vector field.}$$ 
Here we have denoted $x=(x_1, x_2)$ the point of $\R^2.$
After the first retinal process, the visual stimulus reaches the Lateral Geniculate Nucleus without further processing and then the primary visual cortex V1.

\subsection{Simple cells, modeled via contactization and symplectization}
\subsubsection{Odd Simple cells as a contactization of the retinal plane}
Simple cells are the simplest family of cortical cells which process the signal in V1. It is largely accepted that, in presence of a visual stimulus $I:M\to R$, the main functionality of this family of cells is to extract the orientation of level line of $I$ at every point $x =(x_1, x_2)$.  We can always assume that level lines of $I$ are parametrized by arch leth. This means that, if we call $x(s)$ the level line, then $||x'||=1,$ and there exists $\theta$ such that 
\begin{equation}\label{theta_def}
x' = (\cos(\theta), \sin(\theta)).
\end{equation}
The angle $\theta= \theta(s) $ is called orientation at the point $x=x(s)$, and the action of the cell is to detect it. 
In their celebrated paper \cite{[PT99]} Petitot and Tondut modelled the structure of this family of cells as a contact structure, only based on their functionality. 
The model has been generalized in many direction (see for example \cite{[CS06]}, \cite{[SCP08]},  \cite{[SCP09]}), and in the present context it can be expressed as follows. 
From condition \eqref{theta_def}, we deduce  
the relations 
$dx_1 = \cos(\theta) dt$ and $dx_2 = \sin(\theta) dt$.  Removing $dt$ we formally obtain   
$\sin (\theta) dx_1 = \cos(\theta) dx_2$.  Hence, the curve $x=x(s)$ lying on the retinal plane is lifted to a new curve $(x(s), \theta(s))$ in V1, which belongs to the kernel of the $1$-form 
$$\omega_1 = -\sin(\theta)dx_1 + \cos(\theta) dx_2.$$ 

This is a contact form: a contact form is simply defined as an $1$-form such that the volume form is non null:
\begin{equation}\label{contact}\omega_1 \wedge d\omega_1 \not=0. \end{equation}
As a result, the action of the simple cells can be considered as a contactization process, which lift the 2-dimensional complex retinal plane $\C$ to a 3 dimensional contact structure: the space will be obtained by adding the variable $\theta$ to the retinal space 
$$M_1 = \{(x, \theta): x\in \R^2, \theta \in S^1\}. $$
The kernel of the form $\omega_1$ defines the so called horizontal distribution \footnote{recall that a distribution $\Delta$ is a sub-bundle of the tangent space, so that for every $\eta\in M_1$, the space $\Delta_p$ is a sub-space of the tangent space at the point $p$ $T_pM$.}, and a possible choice of generators is defined by the following vector fields $X_{1; i}$, with $i=1, 2$
(as in footnote \ref{footnote1}: the first pedix identifies the family of cells, the second the vector field) :
\begin{equation}\label{campi1}
X_{1;1} = \cos(\theta) \partial_{x_1} + \sin(\theta)\partial_{x_2}, \quad X_{1;2} = \partial_\theta. 
\end{equation}
The fact that $\omega_1$ is a contact form, can be equivalently stated saying that the vector fields satisfy the H\"ormander condition \footnote{a family of vector fields satisfies the H\"ormander condition if the generated Lie algebra coincides with the tangent space at every point.}. If we introduce the metric which makes $X_{1; i}$, with $i=1, 2$ orthonormal, we have defined a sub-Riemannian geometry. 

\subsubsection{Scale dependent simple cells as a symplectization of odd cells}

Starting with a contact structure induced by a $1-$form $\omega_1$ a classical way to define a symplectic structure
is to multiply $\omega_1$ by a real variable and differentiating. This process is called symplectization. 
Precisely the authors of \cite{[SCP08]} and \cite{[SCP09]} 
introduced a new variable $\sigma$ and an enlarged space 
$$  M_2 =  \{ (x,\eta_2): x\in \R^2, \eta_2 = (\theta, \sigma), \theta\in S^1, \sigma \in \R\}.$$
In this space they selected as fundamental 1-form:
\begin{equation*}
 \tilde \omega_1 =e^{-\sigma }\left( -\sin (\theta )dx_1+\cos (\theta
)dx_2\right) , 
\end{equation*}
and took as symplectic form on $T^*M$ the 2-form
 obtained by differentiating $\tilde \omega_1$ with respect to
all its variables. The resulting symplectic 2-form writes:
\begin{eqnarray*}
\omega_2 = d\tilde \omega_1&=&\left( e^{-\sigma }\cos (\theta )dx_1+e^{-\sigma }\sin (\theta
)dx_2\right) \wedge d\theta +\\
&&+ \left( -e^{-\sigma }\sin (\theta )dx_1+e^{-\sigma }\cos (\theta
)dx_2\right) \wedge d\sigma. 
\end{eqnarray*}
A basis of the space is composed by the vector fields 
$X_{2; j}$, where: 
\begin{align}\label{campi2}
& X_{2; 1}= e^{\sigma} \Big(\cos(\theta) \partial_{x_1} + \sin(\theta)\partial_{x_2},\Big), \;\; & X_{2; 2} = \partial_\theta, \\
 & X_{2; 3}  = e^{\sigma} \Big(-\sin(\theta) \partial_{x_1} + \cos(\theta)\partial_{x_2}\Big), \;\;  &X_{2; 4} = e^{\sigma}\partial_\sigma. \nonumber
 \end{align}

\subsection{Complex cells as extension of simple cells space as contact or subriemannian structures}
We have described a process of contactization and symplectization, which, starting from the complex retinal plane, give rise to the 
space of odd and even simple cells. It correctly models the increase of dimensionality of the space, due to the selection of eingrafted variables. We can now go on, further applying a contactization, a simplectization, or building a new sub-Riemannian structure formally adding variables and differential constraints. 
We will see that these extension can describe the functionality of complex cells. 

\bigskip

\subsubsection{Cells sensitive to movement}

A large class of cells in V1 shows a very specific space-time behavior  \cite{DeAngelis1993a}. Many models have been proposed to reproduce these dynamics \cite{DeAngelis1993a,Adelson1985}, \cite{CBS}, and  we show that a contactization of the set of simple cells  can describe the behavior of these families of cells. 
Let us first recall that, due to the aperture problem  \cite{Hildreth} , we can perceive only movement in the direction orthogonal to level lines of the stimulus. 
Using the notation of the previous section this means that the  perceived  velocity is defined as 
$$v = -\sin(\theta) \dot {x} + \cos(\theta) \dot {y} $$
where $\dot {x}$ is the derivative with respect to time variable, 
(and do not coincide with the derivative performed in \eqref{theta_def}). However we are formally in a similar situation, since the new variable $v$ is related to the spatial and time variables via a differential constraint $vdt = \omega_1$. 
This condition naturally suggests to start from $\omega_1$ and extend it by adding  the couple of new variables speed and time: $(v,t)$. The resulting $1$- form is 
\begin{equation} \label{eq:contactform}
\omega_{3}=   \omega_1 - v dt = -\sin(\theta)dx_1 + \cos(\theta) dx_2 - v dt 
. 
\end{equation}
This form is defined on the spatio-temporal phase space with fixed frequency, that is the 5-dimensional manifold
\begin{displaymath}
M_{3} =  \{(x, t,\theta,v): x\in \R^2, t\in \R^+ , \theta \in S^1  v\in  \R \} .
\end{displaymath}
We will also denote $\eta_{3} = (t,\theta,v),$ so that the general element of the space, will be $(x, \eta_3)$. 
We will call horizontal space the kernel of $\omega_{3}$. A basis of this space is constituted of the so-called horizontal or admissible vectors 
\begin{align}\label{eq:vectorfields}
X_{3; 1} =  \cos\theta  \partial_{x_1} + \sin\theta\partial_{x_2},  \;\;
X_{3; 2} = \partial_\theta, \;\;\\ \nonumber
X_{3; 3} = - v \sin(\theta)\partial_{x_1} +  v \cos(\theta) \partial_{x_2} +\partial_t , 
 \;\; 
X_{3; 4} = \partial_v, 
\end{align}
that defines the {\it{horizontal tangent space}}. In particular we note that the Euclidean metric on the horizontal planes makes the vector fields $X_{3; j}$ orthogonal.
Let us explicitly note that 
$[X_{3; 1}, X_{3;2}] =  - \sin(\theta) \partial_{x_1} + \cos(\theta) \partial_{x_2}= [X_{3; 3}, X_{3; 4}]  $
so that the vector fields $X_{3;1}, \cdots, X_{3; 4}$ span the tangent space at every point, and the H\"ormander condition is satisfied. 
The same condition can be expressed recognizing that $\omega_{3}$, is a contact form (see \eqref{contact} for the definition). 

\subsubsection{Cells sensible to frequency and phase}
Let us explicitly note that the same mathematical procedure of starting from the set of simple cells and introducing the time and velocity variables, can be repeated with a couple of different variables satisfying the same differential relation. 
In particular it has been experimentally proved that frequency and phase are selected in the cortex (see \cite{Maffei1977, Levitt1990, Mechler2002, Ribot2013, Tani2012}). The resulting mathematical structure which codes for time and velocity will be the same as coding for frequency and phase: in \cite{[BCS18]} (see also \cite{LM2023}) 
a contactization procedure has been used to extend the form $\omega_4$ to the form 
$$
\omega_{4} = -f sin(\theta) dx_1 + f cos(\theta) dx_2 - ds.
$$
where $f$ is the frequency and $s$ the phase. As before this is a contact form and the horizontal vector fields are formally defined as the generators of the kernel of the one form, and satisfy the H\"ormander condition. 

Let us also mention that the symplectization of this structure give rise to a multiscale phase and frequency model  $M_5$ as in  \cite{[B21]}. 

\subsubsection{Cells selective to curvature}

Finally let us explicitly recall that simplectisation and contactization are not the only way to define a new sub-Riemannian structure. 
Cells sensible to curvature were studied by \cite{AugustZucker}, \cite{Parent} and then  by Petitot in \cite{[P2008]}. Models of the space have been proposed in  \cite{AFCSR,  BC}: the associated set of variables is 
$$M_6= \{(x,\theta, k): x\in \R^2, \theta \in S^1, k\in \R\}.$$
As before we will call $(x, \eta_6)$ the general element of the space, where $\eta_6 = (\theta, k).$ 
The curvature is the derivative of $\theta$ with respect to arch length, so that the associated one form reads 
$$
\omega_6 = d_\theta - k (\cos(\theta) dx_1 + \sin(\theta) dx_2)$$
If we onsider the structure $(M_6, \omega_1, \omega_6)$,  we have obtained a 4 dimensional space, associated to two $1$ forms. Clearly it is neither a contact nor a symplectic space, but it can describe the functionality of cells able to select position, orientation and curvature, by endowing the space with a subriemannian structure, by defining horizontal tangent plane the kernel of the forms $\omega_1$ and $\omega_6$. A set of generators will be: 
\begin{equation}\label{vector6}
X_{6; 1} = \cos(\theta)\partial_{x_1} + \sin(\theta)\partial_{x_2} + k \partial_\theta, \;\; X_{6; 2} = \partial_\theta.
\end{equation}
Note that also these vector fields satisfy the H\"ormander condition.

\subsection{Further extension: hierarchically higher families of cells} From a geometric point of view we can further extend any of these structures, adding new differntial constraints, new $1$ or 2 forms and choosing as horizontal space their kernel. Let us mention for example spaces of jet set \cite{BDo}, stereo \cite{Bo} or structure in the motor cortex \cite{MSC2}.  The element of these spaces will be denoted $(x, \eta_i)$, where $x\in \R^2$ denote the retinal point and $\eta_i$ represent the  eingrafted variable selected by the considered family of cells. We will obtain a family of sub-Riemannian structures $\{M_i: i\in \N\} $, which is not ordered by the inclusion. The families of cells we have described in detail are organized as follows

\begin{figure}[htp]
\centering
\begin{tikzpicture}[
plain/.style={
  draw=none,
  fill=none,
  },
%dot/.style={draw,shape=circle,minimum size=3pt,inner sep=0,fill=black
%  },
net/.style={
  matrix of nodes,
  nodes={
    draw,
    rectangle,
    inner sep=4pt
    },
  nodes in empty cells,
  column sep=0.6cm,
  row sep=-4pt
  },
>=latex
]
\matrix[net] (mat)
{
|[plain]| \parbox{1cm}{\centering Retina } 
          & |[plain]| \parbox{1cm}{\centering Simple cells} 
                       & |[plain]| \parbox{2cm}{\centering Complex cells}  \\
|[plain]| &  |[plain]| &  $\; \; velocity \;\;\atop {time}\; $  \\
|[plain]| &  $orientation$  &  |[plain]|\\
|[plain]| &  |[plain]| & $ frequency\atop {phase} $ \\
 $\; contrast\;$ & |[plain]| & |[plain]| \\
|[plain]| &  |[plain]| &  $ orientation\atop {curvature}$ \\
|[plain]| & $orientation \atop {scale}$  & |[plain]|\\
|[plain]| &|[plain]|  &  $ orientation\atop {curvature, scale}$  \\
};
\foreach \ai in {5}
{\foreach \aii in {3,7}
  \draw[-] (mat-\ai-1) -- node[above] {$\subset$} (mat-\aii-2);
}
\foreach \ai in {3}
 \draw[-] (mat-\ai-2) -- node[above] {$\subset$} (mat-7-2);
 \foreach \ai in {6}
 \draw[-] (mat-\ai-3) --  (mat-8-3);
\foreach \ai in {3}
{\foreach \aii in {2,  4, 6}
  \draw[-] (mat-\ai-2) -- node[above] {$\subset$}(mat-\aii-3);
}
\foreach \ai in {7}
{\foreach \aii in {8}
  \draw[-] (mat-\ai-2) -- node[above] {$\subset$}(mat-\aii-3);
}
\foreach \ai in {2,  4, 6, 8}
\draw[-] (mat-\ai-3) -- node[above]  {other cells}+(3cm,0);
\end{tikzpicture}

\caption{Structure of families of cells. The left column represent the retina, whose cells are sensible to contrast. In the middle column we represent the simple cells, sensible to orientation and scale. Right column: the complex cells process the output of simple cells and select velocity and time, frequency and phase, orientation and curvature.}
\label{fig_m_3}
\end{figure}
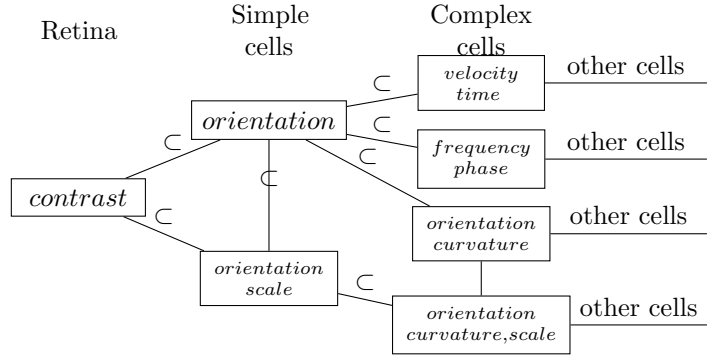

\section{From differential structure to Lie groups representation\\ as models of receptive profiles}

By now we have shown that the formal process of extension of 
sub-Riemannian structure can model families of cortical cells, sensible to different geometric features. 
We have used only the functionality of cells: the selectivity of a cell to a differential feature introduces a differential constraint in the space, which is naturally endowed with a contact, symplectic or sub-Riemannian structure. We would like to show that this formal process is consistent with the shape of cells receptive profiles, and the structure of their set, as they are measured experimentally. 

The \textit{receptive field} (RF) of a visual
neuron is typically modeled as functions, which are positive in the excitatory region, and negative in the inhibitory one. In the examples considered in the previous sections the set of profiles is parametrized by the point in the n-space (for example the 3-space $(x,y,\theta)$ for simple cells): there is a mother profile and the same profile is transported over every point. 

This property will be interpreted as an invariance and described as a Lie group condition. 
{ \it In order to 
recover these profiles, we will  apply the  uncertainty principle induced by the non commutative properties of the vector fields $X_{h; i}$ to identify the mother profile. The profiles will be the minimizers of the uncertainty principle. Then we consider the Lie group associated to the Lie algebra and use the unitary representation of the group to define all the other profiles. 
In other words the group law can be used to generate the whole family of cells starting from the mother profile}. We have adopted this procedure for example in \cite{[BCSS]}.

\subsection{The retinal plane: a commutative group}

In Section \ref{retina} we recognized that the Lie algebra acting on the retinal plane is generated by $\partial_{x_1}$ and $\partial_{x_2}$.  
The associated Lie group is the Euclidean group $\R^2$. This is in good agreement with the well known fact that the
RPs of the retinal ganglion cells are usually modeled by Laplacians of Gaussians  \cite{Marr82},
$$ \psi _{0}(y)=\Delta G(y),$$
where $G(y)= e^{-|y|^{2}}$ is the Gaussian bell and where $\Delta$ is the standard Laplacian. 
This symmetric profile clearly respect the rotational symmetry of the space. The same receptive profile is replicated over each point in the retina, i.e. of the $\R^2$ space 
and this point allows to parametrize the receptive profiles with the point 
where it is centered. 

\subsection{Simple cells Receptive profiles}

\subsubsection{The set of Simple cells Receptive Profiles as a group }

We already recognized that the functionality of simple cells, can be modeled by a contact structure and that the generators of the horizonal distribution are the vector fields $X_{1;1}, X_{1; 2}$, defined in \eqref{campi1}. 
The commutation rules satisfied by the vector fields $X_{1; i}$ univocally identify the group $SE(2)$ of rotation and translation: it coincides with the experimental observation that
the set of simple cells RPs is obtained via translations and 
rotations from a mother profile $\psi
_{1}$ (see \cite{Lee}).

\subsubsection{The mother profile as the minimum of an uncertaintly principle}\label{sec_uncertain1}

The shape of the mother profile has been identified by Daugmann (see \cite{J.G.Daugman}) as a Gabor filter which is minimum of the uncertainty principle in the Heisenberg group. Folland proved in \cite{F89}, that it is possible to state an uncertainty principle any time we have non commuting vector fields, and in \cite{SCfield} we showed that Gabor filters can be  identified as the minima of uncertainty principle associated to the vector fields $X_{1; i}$, and identifies receptive profiles of odd and simple cells, for fixed scale as solution of the following PDE: 
$$X_{1; 1} \psi_1  = i X_{1; 2} \psi_1.$$
The mother profile turn out to be the following one: 
\begin{equation}\label{mother}
\psi _{1}(y)=e^{-|y|^2}e^{2iy_2}.
\end{equation}
Note that this profile can be considered the product of the receptive profile acting on the retinal plane and the complex exponential.  
This shape is in good agreement with the model proposed by Hubel and Wiesel in \cite{[HW77]}, since the profile is obtained by composing profiles of a hierarchically lower family of receptive profile.

\subsubsection{The whole set of profiles and the unitary representation of the group}

The action of the group on $\R^2$ is expressed as follows, 
\begin{eqnarray} \label{changevar}
A_{x,\theta}(y)&=&x + R_\theta y,\;\;
\text{where }\;\;
R_\theta = \left(
\begin{array}{cc}
\cos (\theta ) & -\sin (\theta ) \\
\sin (\theta ) & \cos (\theta )
\end{array}
\right) 
\end{eqnarray}

Since $\theta$ is the first feature extracted we will call 
$\eta_1 = \theta$ and 
the action of the group on the mother profile, 
which generates all the family of filters is expressed as

$$
A_{x, \eta_1}\psi_1(y)=
\psi_1(A^{-1}_{x, \eta_1}(y)).
$$

This can be expressed as a
unitary representation of the group. 
If $G_1$ denotes the $SE(2)$ group, $H$ the Hilbert space $L^2(\R^2),$ a unitary representation 
of the group reads
$$\Phi_1: G_1 \to H\; \quad \Phi_1(x, \eta_1)\psi_1(y) = \psi_1(A^{-1}_{(x, \eta_1)}(y))
.$$
As a result the complete set of filters is obtained as the image of the unitary representation of the group.

\subsubsection{cell sensible to scale}\label{secscale}

In perfect analogy, when we consider the commutation rules associated to the vector fields defined in \eqref{campi2}, identify the group $G_2$ of translation, rotation, and dilation, which indeed is the one experimentally associated to the receptive profiles able to select position, orientation, and scale (see \cite{Lee}). 

It has been proved in \cite{Ali}, that it is not possible to impose the uncertainty principle introduced by Folland and used here in section \ref{sec_uncertain1} to find a minimum. Other uncertaintly criteria which can  provide a minimum, have been introduced in  \cite{Ali} and 
\cite{Sagiv}, but the problem is still partially open. 
Here we choose as a mother profile the  function introduced in \eqref{mother}: $\psi_2 = \psi_1$. 
Then we consider the action of the group $G_2$ on the retinal plane 
\begin{eqnarray*}
A_{x, \eta_2}(y) = A_{x,\theta,\sigma}(y) &=&x +e^{\sigma }R_\theta y
\end{eqnarray*}
As noted in \cite{[SCP08]}, 
 the group action on the mother filter 
provides a family of receptive profiles, 
$A_{x, \eta_2}\psi_2$
which well describes cells sensible to scale.

\subsection{Complex cells Receptive profiles}

\subsubsection{Spatio-temporal Receptive profiles: the Galilean group} When applying this construction to the vector fields $X_{3; i}$ defined in \eqref{eq:vectorfields} we obtain a smooth composition law on $M_3$ which reads as follows 
% \begin{equation}\label{eq:composition}
% \begin{array}{l}
% (\qv,s,\theta,v) \odot (\qp,s',\theta',v')\\
% = (R_\theta (\qp + \textstyle{\binom{v}{0}} s') + \qv, s' + s,\theta' + \theta, v' + v)
% \end{array}
% \end{equation}
\begin{equation}\label{eq:composition}
\begin{array}{l}
(x,t,\theta,v) \odot (x',t',\theta',v')
= (R_\theta (x' + \textstyle{\binom{v}{0}} t') + x, t' + t,\theta' + \theta, v' + v). 
\end{array}
\end{equation}
The manifold $M_3$. together with this composition law can also be identified with a subset $G_3$ of the Galileian group. $G_3$  acts on the space of variables $(y,t) \in \R^2 \times \R^+$ and the action of the group will be denoted 
$A_(x, t, \eta) $. 

An ad hoc version of the uncertainty principle leads to identify as minimum the function 
\begin{equation}\label{mother3}
\psi_3(y,s)= e^{iy_2} e^{-\frac{|y|^2}{\sigma_y^2}-\frac{s^2}{\sigma_s^2}}
\end{equation}
and we will identify all the other filters via the action of $G_3$ 
so that 
the generic receptive profile will be denoted 
$$A_{x, t, \eta_3}\psi_3 (y, s),$$
where we recall that $(x, \eta_3)=(x, \theta, v,t)$. 
Then the whole set of filter will be obtained using the unitary representation 
$$\Phi_3: G_3\to H_3\; \quad \Phi_3(x, \eta_3)\psi_3(y, s) = \psi_3(A^{-1}_{x, \eta_3}(y, s))
,$$
where $H_3= L^2(\R^2 \times \R)$. 
The family of filters obtained in this way is the family of Gaussian Gabor filters centered at position $x \in \R^2$ on the image plane, activated around time $t \in \R$, with spatial frequency $ \cos(\theta), \sin(\theta), \in \R^2$, temporal frequency $\nu \in \R$, spatial width $\sigma_x \in \R^+$ (circular gaussians) and temporal width $\sigma_t \in \R^+$
\begin{equation}\label{eq:insep}
\Phi_3(x, \eta_3)\psi_3(y, s) =  e^{i(R_\theta \cdot (y - x) - \nu (t-s) )} e^{-\frac{|x-y|^2}{\sigma_x^2}-\frac{(t-s)^2}{\sigma_t^2}}
\end{equation}

The structure has been obtained as a complectization of the family of simple cells. 
Correspondingly the mother filter $\psi_3$ is obtained by the by multiplying the mother profile by a function of time:  this seems to be compatible with the model proposed by Hubel and Wiesel \cite{[HW77]}, since receptive profiles are obtained by composition of receptive profiles of hierarchically lower level cells. 
In addition these RP are in good agreement with the experimental evidence provided in  \cite{DeAngelis1993a,Adelson1985} and the model proposed in \cite{CBS}, where  it has been shown that a 3-dimensional sum of Gabor filters can fit very well experimental data of both separable and inseparable RPs.

\subsubsection{Receptive profiles of cells selective for curvature }

The two vector fields introduced in \eqref{vector6},  associated to cells sensitive to curvature, do not commute, and the uncertainty principle stated by Folland, can be expressed as 
$$X_{6,1} \psi_6 = i X_{6,1}\psi_6$$
The solution will be used to model the mother filter for this family of cells. 

From the commutator rules between the vector fields, 
we can deduce that the associated 
Lie group is the Engel one, and that the complete set of receptive fields can be obtained from a fixed one, by the action of the Engel group: 
$$RP_(x, \eta_6)(y) =\Phi(x, \eta_6)\psi_6(y) .$$

\subsection{Receptive profiles of hierarchically higher cells}

The procedure we have described here can be repeated for hierarchically higher families of cells, identified via differential constraints only if the Lie algebra generated by the associated vector fields induces a Lie group structure on the space.

However we can not hope to apply in general Lie groups the Folland uncertainty principle to identify the associated mother  filter. There are indeed groups, in which the system obtained imposing all uncertainty principle has no solutions. It is still an open problem to understand which optimization principles lead to the definition of the mother filter in these groups.  

\section{Cortical Connectivity  and horizontal curves}

\subsection{Fiber bundle structure and the hypercolumnar structure}

The differential structure introduced in section 2 describes the direction of propagation, but all the horizontal directions have the same role. 

This does not happen in the spatial organization of the receptive profiles and in the structure of the connectivity. Experimentally it has been shown that over each retinal point $x$ there is a whole fiber of variables $\eta_i$, extracted by the functionality of the different families of cells and called engrafted variables. The roles of these variables is different, and this organization is called hypercolumnar organization. 

Hoffmann \cite{[HP08]}, and Petitot \cite{[PT99]}, proposed to use the structure of fiber bundle to describe the hypercolumnar. The basis of the space are the variables $x$, and the fiber is constituted by the other variables $\eta_i$. 
However, even the fiber bundle structure alone is not able to capture  the invariance properties of neural space, which are coded in the Lie group.

\subsection{Cortical connectivity of simple cells}

\subsubsection{odd simple cells which select orientation}
We have modelled the set of simple cells as the group $SE(2)$, and denoted $(x, \theta)$ its variables, and we selected a horizontal tangent plane spanned by $X_{1, i}$. The space can also be considered a fiber bundle with basis $B_1= \R^2$, and fiber $F_1 = S^1$. In order to take into account both structures: 
the fiber bundle and the sub-Riemannian structure, 
we remark that the intersection of the basis $B_1$ of the fiber bundle and the horizonal tangent plane $HM_1$ is spanned by the vector $X_{1; 1}$: it has a non trivial projection on the retinal plane. This means that only integral curves of the vector fields with non vanishing component along  $X_{1;1}$ can be considered lifting of curves defined on the retinal planes, and describe meaningfull cortical curves. 

For this reason the structure of connectivity was modelled in \cite{[CS06]} as a curve $\gamma$ satisfying the following condition: 
\begin{equation}\label{lift}\gamma' = X_{1; 1}(x,\theta) +
kX_{1; 2}(x,\theta), \quad \gamma(0) = (x_0,\theta_0). \end{equation}
Here the parameter $k$ varies arbitrarily in $\R$ while the coefficient of $X_1$ is always 1. 
This means that integral curves of $X_{1; 1}$ will have non trivial  projection on the basis of the fiber bundle. 
These vector fields provide a good model of the local association field introduced in \cite{FHH} and \cite{GrossbergMingolla}). 

The model has been modified in \cite{DB} and \cite{BD}, allowing the coefficient of $X_{1; 1}$ to vanish,  considering geodesics of the subriemannian space and discarding the fiber bundle structure. The curves better fit the association fields. However the authors proved that,  even if they start with this apparently more general condition, they are obliged to stop the geodesics when the coefficient of $X_{1; 1}$ vanish, since the curves produces cusps. 
A more natural approach has been introduced in \cite{MSC1} where the authors go back to the original model and look for a curve of minimal lenght directly in the set of curves expressed in the form \eqref{lift}. the curves coded in the cortex, and whose projection does not produce cusps. 
%
%Since the vector field $X_{1; 1}$ has a role completely different from $X_{1; 2}$, in \cite{} it has proposed to model the connectivity kernel can be modelled as 
%the fundamental solution of the Fokker Plank equation 
%$$X_{1; 1} + X_{1; 2}^2 =0.$$
%The kernel exibits a totally anysotropic behavior, which correctly models the cortical connectivity. 
%{\color{red}qui bisogna citare qualcuno?}

\subsubsection{Simple cells sensible to scale}

Simple cells sensible to scale have been modelled as a symplectic structure in Section \ref{secscale}. A symplectic space is a quasi complex space, where the  multiplication for $i$ is replaced by the linear transform 
$J$ such that $$J( X_{2; 1}) =  X_{2; 2},  \; J(  X_{2; 2}) = -  X_{2; 1},  \; J( X_{2; 3}) =   X_{2; 4},  \; J( X_{2; 4}) = -  X_{2; 3}. $$
Hence in \cite{[SCP09]} together with J. Petitot, the authors first considered separately integral curves of vector fields $X_{2; 1}$ and  $X_{2; 2}$
and of the vector fields $X_{2; 3}$ and  $X_{2; 4}$ . 
Their pattern generated by the first one is in good agreement with the pattern
of long range connections found both in neurophysiological and psychophys-
ical experiments. 
Integral curves of the vector fields $X_{2; 3}$ and  $X_{2; 4}$ 
model the trans-axial connections, which extends orthogonally from the orientation
axis of the cell.

\subsection{Cortical connectivity of complex cells} 

\subsubsection{cells sensible to movement}
The set of cells sensible to movement has been modelled as the sub-Riemannian structure defined by the manifold $M_3=\{(x, t, \eta_3)$, 
where $\eta_3 =(\theta, v)$, and the distribution generated by the vector fields $X_{3, i}$ defined in \eqref{eq:vectorfields}.  The space can also be considered a fiber bundle with basis $B_3= \R^2 \times \R^+ $,  of variables $(x,t)$ and fiber $F_3 = S^1\times \R$, which contains the eingrafted variables $\eta_3 =(\theta, v)$. We select the vector fields with non trivial projection on the retinal plane: $X_{3; 1}$ and $X_{3; 3}$. As a result, the connectivity will be modelled by the integral cuves of the vector fields $X_{3; 1}+ k X_{3; 2}$, which describe long range connectivity for boundary completion as in the odd simple cells. In addition integral curves of the vector fields $X_{3; 3}+ k X_{3; 2}$ related to the perception of movement. 

\subsection{Cortical connectivity of hyerarchically higer cells} 
All families of cells considered up to now have been described as manifolds $M_h$ 
parametrized as $(x,\eta_h)$, where 
$x$ is the retinal variables and $\eta_h$ describes the eingrafted variables. Hence they have a natural structure of fiber bundle, where $x$ belong to the basis and $\eta$ to the fiber. For fixed $h$ we selected vector fields  $X_{h; 1}, X_{h; 2}, ...$ which define a subriemannian metric. The connectivity is modelled by integral curves of linear combination of these vector fields, with coefficient 1 for the vector field with non trivial projection on the retinal plane.

\section{Action and functionality of receptive profiles, as Bargmann transform}

We started from an abstract notion of functionality as selection of differential features to express our model. We obtained contact, symplectic and subriemannian structures, directly from the differential properties of the features. Up to now,
the model seems expressed in terms of universal and geometrical founding principles. The  fit between the properties of Lie group and the shape of minima of uncertainty principle from one side and the structure and the shape of receptive profiles, provides a first validation of the project. We verified that integral curves compatibles with the horizontal distribution and the fiber bundle structure are a good model for the cortical connectivity. Finally  we provide an other validation of the model, proving that receptive profiles obtained in this way, are able to select the desired features. 

\subsection{Action of receptive profiles as Bargmann transform}

The overall output $O$ of the parallel filtering performed by receptive profiles is given by the
integral of the signal $I$ times the bank of filters:
\begin{eqnarray}\label{output}
O(x, t, \eta) =\int_{M}RP_{x, t, \eta}(y) I(y) dy
.
\end{eqnarray}
We present here an unitary description which applies to all the considered families of cells considered before. In particular for $t=0$ we recover family of cells independent of time. 
As remarked in \cite{SCfield}
the whole output will be identified with
a Bargmann transform:
\begin{equation}\label{bargman}
B(I)(x, t, \eta) =  \int_{M} \Phi(x, t, \eta)\psi_h(y) I(y) dy.  \end{equation}

This transform maps the image $I$ defined on the 2D space, to a
function which depends on all the variables of the total space of the cortical layer. 
This integral has the natural meaning of projection of $I$ in the direction
of the specific element $\Phi(x, t, \eta)$ of the frame.

\subsection{Action of intracortical connectivity and feature selection }

Several models have been presented to explain the selectivity of features 
in the primary visual cortex. Even if
the basic mechanism producing strong orientation selectivity is
controversial (see for example \cite{MKP, NSS, SWMcLS}), nevertheless it is evident that the intra-cortical circuitry
is able to filter out all the spurious directions and to strictly
keep the direction of maximum response of the simple cells. 

\subsubsection{Non maximal suppression and orientation selectivity in simple cells}

While considering simple cells sensible to orientation, the output   $O$ in formula \eqref{output} 
is a function of the variables $x$ and $\theta$ alone. 
Since the output $O$ is a regular function, its maximum at a value $\theta^* = \theta^*(x)$
\begin{displaymath}
\max_{\theta} O (x,\theta) = O (x,\theta^*) 
\end{displaymath}
This mechanism was introduced by \cite{[CS06]}, 
who proved that the vector $(- \sin\theta^*, \cos\theta^*)$  is 
orthogonal to the level lines of $I(x)$. 
As a consequence, the level lines of $I$ are lifted to curves admissible in the sense that their tangent vector lies in the kernel of the form $\omega_1$. And this provides a validation of the model since it ensures that the functionality of these filters is orientation selection. 

\subsubsection{Non maximal suppression in complex cells and selectivity of time and movement}

While taking into account time, the output $O$ defined in formula \eqref{output} will be a function of variables $(x, t, \eta_3)$ defined on $M_3$.  
The output $O$ takes its maximum at the values $\theta^* $ and $v^* $ 
\begin{displaymath}
\max_{\theta , v} O (x, t,\theta,v) = O (x, t,\theta^*,v^*). 
\end{displaymath}
In perfect analogy with the lower dimensional case,  
the vector $(- \sin\theta^*, \cos\theta^*, v^*)$ is orthogonal to the level set of $I$. The vector $(- \sin\theta^*, \cos\theta^*)$ is orthogonal to the spatial level line, and the scalar $v^*$ represents the apparent velocity in this direction (see \cite{CBS} for the proof). 

In this way the receptive profiles are able to select the features of orientation and velocity. 

\subsubsection{Hierarchically higher level cells}

For general families of cells, independent of time and selective to a feature $\eta$, 
the output $O$ will be a function of the variables $(x,\eta)$ and the non maxima suppression mechanism will be expressed as the existence of a value $\eta^*$
\begin{displaymath}
\max_{\eta} O (x,\eta) = O (x,\eta^*) 
\end{displaymath}
The value of $\eta^*$ represents the selection of the value of the engrafted variables performed by the layer of cells.

\section{Discussion}

We presented a model of the visual cortex able to describe different families of cells, and based on three genearal principles.  The functionality of each family of cells defines a one form. The invariance properties of the set of cells are captured by a Lie group structure. The hypercolumnar structure is modelled by a fiber bundle structure. 

Further models of contact/symplectic structures in neurogeometry have been proposed recently to model the functional architecture of M1 cells encoding hand movement direction
\cite{MSC2}  and to model the functional architecture of stereo vision \cite{Bo}. Metric models of the visual cortex have been presented in  \cite{MCS}\cite{MSC}. 

Other models have been proposed to face the problem of the sparseness of orientation hypercolumns in the visual cortex. In fact, being the dimension of a hypercolumn about $1 mm^2$ and being a typical dimension of V1 in primates around $12-16 cm^2$, the number of hypercolumns will be respectively of $30\times40 - 40\times40$, that is absolutely insufficient to sample a high resolution stimulus image, even taking into account the anisotropy of sampling. To overcome the problem, it has been proposed that all the cells of a  hypercolumn are actively sampling the image stimulus and contribute to its reconstruction. A heterogeneous Poisson problem has been posed and solved in \cite{SCP22}\cite{SGC23}, starting from the concept of heterogenous differential operator proposed in \cite{SCP19}.

Finally a new line of research emerged from the integration of geometric technics and learning procedures \cite{BCS} \cite{BMSC}.

%
%\begin{itemize}
%\item{simmetrie nelle architetture funzionali: quanto la modellizzazione che 
%introduce le simmetrie di Lie modella correttamente i dati. 
%Le corteccie primarie sono molto molto vicine al dato sensibile e quindi 
%non e' riduttivo che ne ereditino le simmetrie. 
%Citare: nuclei di correlazione e misure di co-occorrenza
%}
%\item{Presenza di effetti legati alle pregnanze: in genere ogni processo di learning e' sempre modulato da meccanismi di rinforzo legati alla fisiologia dell'organismo. Questi sono particolarmente presenti nelle corteccie superiori, e possono essere trascurati qui, per la prevalenza delle salienze rispetto alle pregnanze
%}
%
%\item{esistenza del fibrato posizione / engrafted variables. 
%C'e' un problema di discretizzazione della base del fibrato. I pinwheel sono troppo grandi per rappresentare in modo fine. 
%}
%
%\item{Connessione fra cellule eterogenee. Abbiamo considerato considerare la connettivita' all'interno della stessa famiglia di cellule, mentre sarebbe interessante vedere come si connettono cellule di tipo diverso.  Problema di accoppiamento di cellule eterogenee fra layer diversi, e nello stesso layer
%}
%\item{Problema della ricostruzione dell'immagine in corteccia V1: con la discretizzazione a Pinwheels non si riesce. Con Davide non e' su base neurale. E il lavoro nostro. 
%}
%\end{itemize}

%
%Da citare
%Biol. Cybern. 54, 107-114 (1986) Biological
%Cybernetics
%  Springer-Verlag 1986
%A Mathematical Model of the Primary Visual Cortex and Hypercolumn
%K. Okajima 

\end{document}